\documentclass[aps,superscriptaddress,prl,twocolumn,nofootinbib]{revtex4}

\usepackage{graphicx}
\usepackage{epstopdf}
\usepackage{subfigure}
\usepackage{epsfig}
\usepackage{amsmath,amssymb,amsfonts,bm}
\usepackage{color}
\usepackage[utf8]{inputenc}
\usepackage{color}
\usepackage[bookmarks,
                bookmarksopen = true,
                bookmarksnumbered = true,
                linktocpage,
                colorlinks = true,
                linkcolor = blue,
                urlcolor  = blue,
                citecolor = blue,
                anchorcolor = green,
                hyperindex = true,
                hyperfigures]
                {hyperref}

\usepackage[normalem]{ulem}
\renewcommand{\sout}{\bgroup \color{red} \ULdepth=-.5ex \ULset}

\begin{document}

\title{Charmed hadron chemistry in relativistic heavy-ion collisions}

\author{Shanshan Cao}
\affiliation{Cyclotron Institute and Department of Physics and Astronomy, Texas A\&M University, College Station, TX, 77843, USA}
\affiliation{Department of Physics and Astronomy, Wayne State University, Detroit, MI, 48201, USA}
\author{Kai-Jia Sun}
\affiliation{Cyclotron Institute and Department of Physics and Astronomy, Texas A\&M University, College Station, TX, 77843, USA}
\author{Shu-Qing Li}
\affiliation{Department of Physics and Information Engineering, Jining University, Qufu, Shandong, 273155, China}
\affiliation{Institute of Particle Physics and Key Laboratory of Quark and Lepton Physics (MOE), Central China Normal University, Wuhan, Hubei, 430079, China}
\author{Shuai Y.F. Liu}
\affiliation{Quark Matter Research Center, Institute of Modern Physics, Chinese Academy of Sciences, Lanzhou, Gansu, 073000, China}
\affiliation{University of Chinese Academy of Sciences, Beijing, 100049, China}
\affiliation{Cyclotron Institute and Department of Physics and Astronomy, Texas A\&M University, College Station, TX, 77843, USA}
\author{Wen-Jing Xing}
\affiliation{Institute of Particle Physics and Key Laboratory of Quark and Lepton Physics (MOE), Central China Normal University, Wuhan, Hubei, 430079, China}
\author{Guang-You Qin}
\affiliation{Institute of Particle Physics and Key Laboratory of Quark and Lepton Physics (MOE), Central China Normal University, Wuhan, Hubei, 430079, China}
\affiliation{Nuclear Science Division, Lawrence Berkeley National Laboratory, Berkeley, CA, 94270, USA}
\author{Che Ming Ko}
\affiliation{Cyclotron Institute and Department of Physics and Astronomy, Texas A\&M University, College Station, TX, 77843, USA}

\date{\today}

%%%%%%%%%%%%%%%%%%%%%%%%%%%%%%%%%%%%%%%%%%%%%%%%%%%%%%%%%%%%%%%%%%%%

\begin{abstract}
We develop for charmed hadron production in relativistic heavy-ion collisions a comprehensive coalescence model that includes an extensive set of $s$ and $p$-wave hadronic states as well as the strict energy-momentum conservation, which ensures the boost invariance of the coalescence probability and the thermal limit of the produced hadron spectrum. By combining our hadronization scheme with an advanced Langevin-hydrodynamics model that incorporates both elastic and inelastic energy loss of heavy quarks inside the dynamical quark-gluon plasma, we obtain a successful description of the $p_\mathrm{T}$-integrated and differential $\Lambda_c/D^0$ and $D_s/D^0$ ratios measured at RHIC and the LHC. We find that including the effect of radial flow of the medium is essential for describing the enhanced $\Lambda_c/D^0$ ratio observed in relativistic heavy-ion collisions. We also find that the puzzling larger $\Lambda_c/D^0$ ratio observed in Au+Au collisions at RHIC than in Pb+Pb collisions at the LHC is due to the interplay between the effects of the QGP radial flow and the charm quark transverse momentum spectrum at hadronization. Our study further suggests that charmed hadrons have larger sizes in medium than in vacuum. 
\end{abstract}

\maketitle

%%%%%%%%%%%%%%%%%%%%%%%%%%%%%%%%%%%%%%%%%%%%%%%%%%%%%%%%%%%%%%%%%%%%%

{\em Introduction. --}
Relativistic heavy-ion collisions provide a unique opportunity to study the properties of the color deconfined state of nuclear matter, known as the Quark-Gluon Plasma (QGP)~\cite{Shuryak:2014zxa}. Heavy quarks are considered as a clean probe of the QGP properties since they are mostly produced in the early stage of nuclear collisions due to the negligible thermal production as a result of their large masses~\cite{Zhang:2007dm,Liu:2016zle,Zhou:2016wbo}. Extensive studies have been devoted to the production, evolution and nuclear modification of heavy quarks in relativistic nuclear collisions~\cite{Rapp:2018qla,Cao:2018ews,Xu:2018gux,Dong:2019byy}. Recent experiments at the Relativistic Heavy-Ion Collider (RHIC) and the Large Hadron Collider (LHC) have shown a non-monotonic transverse momentum ($p_\mathrm{T}$) dependence of the $D$ meson nuclear modification factor ($R_\mathrm{AA}$) at low $p_\mathrm{T}$~\cite{Adam:2018inb} and a large charmed baryon-to-meson ratio~\cite{Adam:2019hpq,Acharya:2018ckj,Vermunt:2019ecg}. In particular, the observation of a larger $\Lambda_c/D^0$ ratio in Au+Au collisions at RHIC compared to that in Pb+Pb collisions at the LHC is an unexplained puzzle in the field of relativistic heavy-ion collisions. 

To understand the above intriguing experimental results, we develop a comprehensive model for the production of heavy quarks and their hadronization to heavy flavor hadrons in high energy collisions. In these collisions, light hadrons of low $p_\mathrm{T}$ are known to mainly originate from the thermal emission of the produced fireball~\cite{Adler:2003kt,Adams:2003am,Abelev:2014pua}, while high-$p_\mathrm{T}$ light and heavy flavor hadrons and jets are dominantly produced from the fragmentation of hard partonic jets~\cite{Abelev:2013fn,Qin:2015srf,Kumar:2019bvr,Xing:2019xae}. For hadrons of intermediate $p_\mathrm{T}$, quark coalescence is believed to be the dominant mechanism for their production~\cite{Greco:2003xt,Fries:2003vb,Hwa:2002tu}. Various coalescence approaches have been developed in the past. For models with simplified assumptions, such as the equal-velocity coalescence~\cite{Song:2018tpv} or the coalescence between neighboring quarks~\cite{Zheng:2019alz}, although they can describe some features of the charmed hadron chemistry, they lack quantitative calculations of the microscopic coalescence probabilities as functions of the quark distance in the phase space. To improve this shortcoming, a resonant recombination model has been developed in Refs.~\cite{He:2011qa,He:2019vgs} by connecting the coalescence probability with the resonant scattering rate of heavy quarks in the QGP. There is also a coalescence model that is based on the wave function projection and thus contains more detailed information about the hadron structure as compared to other approaches. Such an approach was first introduced for understanding light nuclei production from the coalescence of nucleons~\cite{Butler:1961pr,Sato:1981ez,Csernai:1986qf,Dover:1991zn} and later extensively applied to study the production of charmed hadrons and hadrons from jets from the coalescence of constituent quarks~\cite{Oh:2009zj,Gossiaux:2009mk,Song:2015sfa,Cao:2016gvr,Plumari:2017ntm,Li:2018izm,Zhao:2018jlw,Cho:2019lxb,Han:2016uhh}.

In this Letter, we develop a comprehensive coalescence model that not only includes an extensive set of $s$ and $p$-wave hadronic states but also imposes the strict energy-momentum conservation. The inclusion of $p$-wave states allows a proper normalization of the total coalescence probability for zero-momentum charm quarks using reasonable values of in-medium hadron sizes, which turns out to be important for understanding the observed $\Lambda_c/D^0$ ratio. The rigorous 4-momentum conservation guarantees the boost invariance of the coalescence probability and the thermal limit of the produced hadron spectra. By combining our new hadronization approach with the Langevin-hydrodynamics model~\cite{Cao:2013ita,Cao:2015hia} that simulates elastic and inelastic energy loss of heavy quarks in a realistic QGP medium, we provide the first simultaneous description of the chemical compositions of charmed hadrons measured at both RHIC and LHC. 
%We find that the larger $\Lambda_c/D^0$ ratio observed in Au+Au collisions at RHIC than in Pb+Pb collisions at the LHC can be naturally explained in terms of the interplay between the collective flow of produced QGP and the charm quark spectra.

{\em The coalescence model. --}
The momentum distribution of hadrons produced from quark coalescence is given by
\begin{equation}
\label{eq:distributionH}
f_h(\bm{p}_h')=\int \Big[\prod_{i} d\bm{p}_i f_i (\bm{p}_i)\Big] W(\{\bm{p}_i\})\delta (\bm{p}_h'-\sum_{i} \bm{p}_i),
\end{equation}
where $\bm{p}_h'$ is the 3-momentum of the hadron, and $\bm{p}_i$ is that of each constituent quark with $i$ running from 1 to 2 (3) for the produced meson (baryon). The Wigner function $W$ represents the probability for quarks to coalesce into the hadron and is calculated from the wave function overlap between the free quark states and the hadron bound state.

Assuming the quark wave functions in the meson to be those of a harmonic oscillator potential, the Wigner functions for mesons in the $s$ and $p$-wave states are then given, respectively, by
%~\cite{Baltz:1995tv,Chen:2006vc}:
\begin{align}
\label{eq:WignerS}
&W_s=g_h\frac{(2\sqrt{\pi}\sigma)^3}{V} e^{-\sigma^2\bm{k}^2},\\
\label{eq:WignerP}
&W_p=g_h\frac{(2\sqrt{\pi}\sigma)^3}{V} \frac{2}{3}\sigma^2\bm{k}^2 e^{-\sigma^2\bm{k}^2}.
\end{align}
In the above, the spatial part of the Wigner function has been averaged over the volume $V$, which will be cancelled by the volume factor associated with the momentum distribution functions $f_i$ of light quarks; $g_h$ is the statistical factor for the spin-color degrees of freedom; $\bm{k}$ is the relative momentum between the two constituent quarks defined in their center-of-momentum frame (the meson rest frame); $\sigma=1/\sqrt{\mu\omega}$ with $\mu$ being the reduced mass of the quark and antiquark in the meson, and $\omega$ is the oscillator frequency, which can be directly related to the meson radius~\cite{Oh:2009zj}. The constituent quark masses at the hadronization temperature $T_\mathrm{c}$ are taken as $m_{u,d}=0.3$~GeV for $u$ and $d$ quarks and $m_s=0.4$~GeV for $s$ quark. For the charm quark mass, we take its value to be $m_c=1.8$~GeV as suggested by the $T$-matrix approach that takes into account the effect due to the confining potential~\cite{Riek:2010fk,Liu:2017qah}. This coalescence model can be extended to baryon production by first combining two quarks and then combining their center-of-momentum with the third quark. We symmetrize all possible internal configurations of the baryon by allowing any two of the three quarks to combine first to carry non-zero orbital angular momentum in the case of baryon formation in the $p$-wave state. 

In contrast to previous works~\cite{Oh:2009zj,Plumari:2017ntm,Cho:2019lxb}, we include in the present study all $s$ and $p$-wave hadron states allowed by the spin-orbit coupling in a full 3-dimensional calculation, which covers nearly all charmed hadron species reported in the PDG~\cite{Tanabashi:2018oca}. Take $D^0$ ($c\bar{u}$) for instance, different quark spin combinations ($S = 0, 1$) allow the $s$-wave (orbital angular momentum $L=0$) to construct the states of the total angular momentum $J=0$ ($D^{0}$) and $J=1$ ($D^{*0}$). For $p$ wave ($L=1$), $S=0$ gives $J=1$ ($D_1^0$), and $S=1$ allows $J=0$ ($D_0^{*0}$), $J=1$ ($D_1^{*0}$) and $J=2$ ($D_2^{*0}$). Similar construction is also applied to three-quark states for baryon formation. To compare to experimental data, we allow the decay of all $D^{*0}$'s to $D^0$'s, 68\% $D^{*+}$'s to $D^0$'s and 32\% $D^{*+}$'s to $D^+$'s. For other excited states of $D^0$ and $D^+$ whose decay branching ratios are not clearly measured, we assume them to be 50\% to $D^0$ and 50\% to $D^+$. All excited states of $D_s$ are decayed to $D_s$. We also let excited states of $\Lambda_c$ and all states of $\Sigma_c$ decay to $\Lambda_c$. We only consider the coalescence between a single charm quark with thermal light quarks ($u$, $d$ and $s$) in this work. Formation of multi-charmed hadrons is neglected and may contribute to a small correction to our result.

\begin{figure}[tb]
  \epsfig{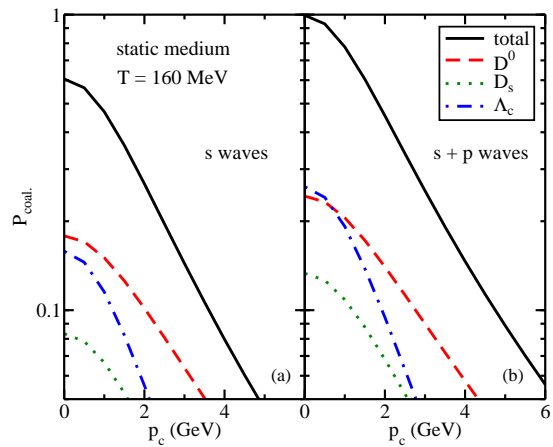}
  \caption{(Color online) Coalescence probabilities of a charm quark in a static medium of temperature 160 MeV for forming different charmed hadrons as functions of its momentum.}
  \label{fig:plot-Pcoal}
\end{figure}

In Fig.~\ref{fig:plot-Pcoal}, we present the coalescence probabilities of charm quarks in a static medium of temperature $T=160$~MeV for different charmed hadron species as a function of the charm quark momentum $p_c$. These probabilities are obtained from integrating the hadron spectrum in Eq.~(\ref{eq:distributionH}) by taking the charm quark distribution as $\delta({\bm{p}_i-\bm{p}_c})$ and the light quark distribution as $g_i V / (e^{E_i/T}+1) /(2\pi)^3$ with $E_i$ and $g_i$ being the light quark energy and degeneracy, respectively. In obtaining these results, we also include the contribution from thermal gluons by converting a pair of thermal gluons, whose masses are taken to be $m_g=0.3$~GeV,  into a quark-anti-quark pair $gg\rightarrow q\bar{q}$ with equal distributions among $u$, $d$ and $s$ flavors~\cite{Plumari:2017ntm,Cho:2019lxb}. For the oscillator frequency, its value is taken to be $\omega=0.24$~GeV so that the total coalescence probability for a zero-momentum charm quark, which does not fragment to hadrons, is equal to one when all $s$ and $p$-wave charmed mesons and baryons are included in the calculation. This value of $\omega$ corresponds to an in-medium radius of $r=\sqrt{3/(2\mu\omega)}=0.97$~fm for $D^0$, which is larger than its value $r=0.83$~fm in vacuum when $\omega=0.33$~GeV is used~\cite{Oh:2009zj,Hwang:2001th}. In principle, different oscillator frequencies can be used for different hadron species. In the absence of theoretical studies of the in-medium sizes of charmed hadrons, we assume for simplicity the same oscillator constant for all charmed hadrons. We believe that our approach of tuning the single model parameter $\omega$ to normalize the coalescence probability is more physically motivated than that in Refs.~\cite{Plumari:2017ntm,He:2019vgs} by multiplying an arbitrary overall normalization factor to the Wigner function. Assuming the same oscillator frequency $\omega$ does not mean the same sizes for all charmed hadrons. Here we present the radius of a two-body system calculated from its wavefunction. In literature, one may define the charge radius (or mass radius) of a hadron by weighing this wavefunction radius with the charge (or mass) of constituent quarks of a hadron~\cite{Oh:2009zj,Sun:2017ooe}. With the same $\omega$, $\Lambda_c$ has similar charge radius, but larger mass radius, compared to $D^+$. As shown in Fig.~\ref{fig:plot-Pcoal}, including the contribution from $p$-wave charmed hadrons significantly enhance the total coalescence probability of a charm quark. 
%Since this enhancement is stronger for $\Lambda_c$ baryons than $D^0$ and $D_s$ mesons, the inclusion of the $p$-wave contribution can also increase the baryon-to-meson ($\Lambda_c/D^0$) ratio from our model. 
%We note that only the contribution from the $s$-wave charmed hadrons have been included in previous studies~\cite{Oh:2009zj,Plumari:2017ntm,Cho:2019lxb}, where \sout{only the $s$-wave contribution is considered and} the calculations are \com{further} carried out \sout{on} \com{for} the 2-dimensional transverse plane (for cylindrical medium).

With the momentum-dependent charm quark coalescence probabilities shown in Fig.~\ref{fig:plot-Pcoal}(b), the hadronization of charm quarks on the QGP boundary ($T_\mathrm{c}=160$~MeV) can be treated by first boosting each charm quark to the rest frame of the local fluid cell within our Langevin-hydrodynamics simulation~\cite{Cao:2013ita,Cao:2015hia} and then determining the production probability of a selected charmed hadron from these coalescence probabilities. The momentum of this charmed hadron is determined from the differential spectrum given in Eq.~(\ref{eq:distributionH}) and then boosted back to the global center-of-mass frame of nuclear collisions. Based on the total coalescence probability given in Fig.~\ref{fig:plot-Pcoal}(b), charm quarks that do not coalesce with thermal quarks are fragmented to hadrons via \textsc{Pythia} simulation~\cite{Sjostrand:2006za}, in which the default Peterson fragmentation function is utilized. This is referred to as the fragmentation-coalescence approach.

\begin{figure}[tb]
  \epsfig{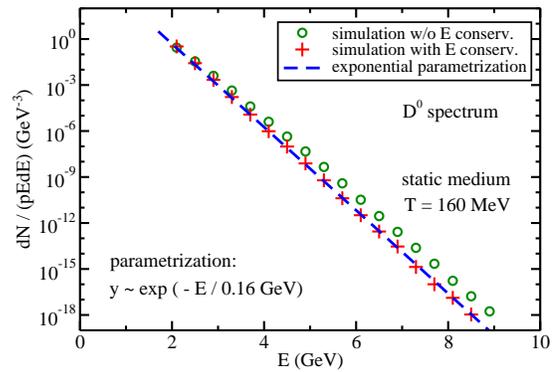}
  \caption{(Color online) Energy spectra of $D^0$ mesons produced from the coalescence of thermal charm quarks in a static medium of temperature $160$~MeV with crosses and circles denoting, respectively, results obtained with and without the 4-momentum conservation. The dashed line is a parametrization of the thermal distribution.}
  \label{fig:plot-check-thermal}
\end{figure}

%{\em Energy conservation and the thermal limit. --}
The long-standing problem in the coalescence model related to the lack of the energy conservation, which breaks the boost invariance of the coalescence probability, is overcome in the current work by equating the total energy of the combined quarks to that of the produced hadron ($E_h'=\sum_i E_i$) and then letting it decay into an on-shell charmed hadron and a pion. Because of the kinetic energies of coalescing quarks, we rarely encounter cases where such decay is not energetically feasible, and they are ignored when this happens. The final energy and momentum ($E_h$, $\bm{p}_h$) of each charmed hadron is determined by the corresponding 2-body decay process based on its initial energy and momentum ($E_h'$, $\bm{p}_h'$) from Eq.~(\ref{eq:distributionH}). This setup guarantees the 4-momentum conservation and preserves the boost invariance.

Figure~\ref{fig:plot-check-thermal} shows the $D^0$ energy spectrum obtained from calculations using the Maxwell-Boltzmann distribution at $T = 160$ MeV, i.e., $e^{-E/0.16~{\rm GeV}}$, for both charm and light quarks energy spectra. It is clearly seen that the $D^0$ energy spectrum from our improved coalescence approach (pluses) follows a Maxwell-Boltzmann distribution at same temperature (blue dashed line). This is in contrast with the result from earlier models without energy conservation (circles), which only follows the shifted thermal distribution at this temperature in the high energy region (above 6~GeV), where the hadron binding energy can be neglected, but shows deviations at low energies. We have verified that the same conclusion holds for $D_s$ and $\Lambda_c$. Note that while the energy-momentum conservation improves the thermal equilibrium limit of the coalescence model, i.e., the energy dependence of the Maxwell-Boltzmann distribution, the chemical equilibrium limit is not required by the sudden approximation of the model. Thus, there exists a normalization factor of the hadron yield when comparing the exponential parametrization with the charmed hadron spectra in Fig.~\ref{fig:plot-check-thermal}.

%\begin{figure}[tb]
%  \epsfig{file=plot-RAA_HM_e.eps, width=0.4\textwidth, clip=}
%  \caption{(Color online) $R_\mathrm{AA}$'s of $D$ mesons (a) and $D$-decayed electrons (b). Results for different hadronization mechanisms are compared. Data are taken from Refs.~\cite{Adamczyk:2014uip,Adam:2018inb,Oh:2017usm}.}
%  \label{fig:plot-RAA_HM_e}
%\end{figure}

%{\em Nuclear modification of charmed hadrons. --}

%In Fig.~\ref{fig:plot-RAA_HM_e}, we present the nuclear modification factor $R_\mathrm{AA}$'s of $D^0$ mesons and $D$-decayed electrons. The diffusion coefficient is set as $D_\mathrm{s}(2\pi T)=3$ to describe the high-$p_\mathrm{T}$ data where $R_\mathrm{AA}$ is dominated by the in-medium energy loss of heavy quarks and the uncertainties from the hadronization process is negligible. Our result shows that fragmentation alone is not sufficient to describe the non-monotonic $p_\mathrm{T}$ dependence of the $D$-meson $R_\mathrm{AA}$ at low $p_\mathrm{T}$. The inclusion of coalescence is necessary to yield the bump structure of $R_\mathrm{AA}$ between 1-3~GeV observed in the experimental data. Note that the magnitude of the coalescence enhancement depends on the strength of the QGP radial flow, as we will see later.

\begin{figure}[tb]
  \epsfig{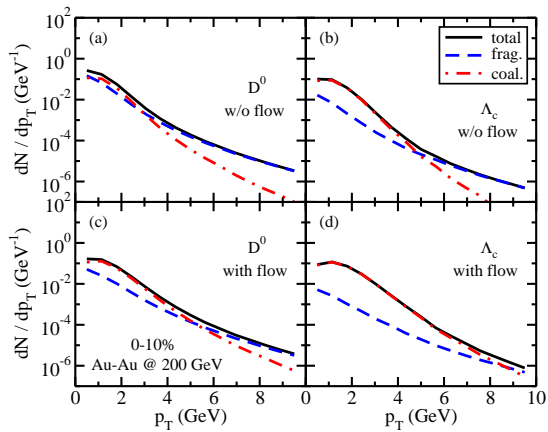}
  \caption{(Color online) Transverse momentum spectra of $D^0$ and $\Lambda_c$ in 0-10\% Au-Au collisions at 200 GeV (normalized to one charm quark) from different hadronization processes: coalescence (dash-dotted), fragmentation (dashed), and both (solid). Results with (lower panel) and without (upper panel) the QGP flow effects are compared.}
  \label{fig:plot-spectra-AuAu200_0-10}
\end{figure}

{\em Charmed hadron chemistry. --}
For charm quark production in relativistic heavy-ion collisions, we employ the advanced Langevin-hydrodynamic model developed in Refs.~\cite{Cao:2013ita,Cao:2015hia}. In this work, the momentum distribution of heavy quarks produced in the initial hard scatterings is obtained from the FONLL calculation~\cite{Cacciari:2001td,Cacciari:2012ny,Cacciari:2015fta} using the parton distribution function CT14NLO~\cite{Dulat:2015mca} for free proton and EPPS16~\cite{Eskola:2016oht} for nuclei. In FONLL, we use the current charm quark mass 1.3~GeV for their production, and set both renormalization and factorization scales as the charm quark transverse mass. The central values of the EPPS16 parametrization are adopted for the nuclear shadowing effect. The initial spatial distributions of both heavy quarks and the entropy density of the QGP medium in heavy-ion collisions are obtained from the Monte-Carlo-Glauber model. An improved Langevin model, which takes into account both elastic and inelastic energy loss of heavy quarks propagating through a (2+1)-dimensional viscous hydrodynamic medium~\cite{Song:2007fn,Song:2007ux,Qiu:2011hf}, is then used to calculate the nuclear modification of heavy quarks. This Langevin approach has only one free parameter, the spatial diffusion coefficient $D_\mathrm{s}$ by convention. A good description of the $R_\mathrm{AA}$ of both heavy flavor mesons and their decayed leptons can be obtained with the values $D_\mathrm{s}(2\pi T)=3.5$ for RHIC and $4$ for the LHC~\cite{Li:2020kax}. The systematic uncertainties in various model components, such as the initial charm quark spectra and the temperature dependence of the diffusion coefficient, have been discussed in detail in Ref.~\cite{Li:2020kax}.

Charm quarks from the Langevin-hydrodynamic model at the $T_\mathrm{c}=160$~MeV hypersurface of the QGP are converted into different species of charmed hadrons using our new fragmentation-coalescence approach. As shown in Fig.~\ref{fig:plot-spectra-AuAu200_0-10}, the hadron chemistry is sensitive to the collective flow of the QGP, i.e., heavier hadrons experience a stronger boost (or gain more transverse momentum) from the medium flow than lighter hadrons. These can be clearly seen by comparing the $D^0$ and $\Lambda_c$ spectra from full simulations in central Au-Au collisions at RHIC between with (lower panels) and without (upper panels) the presence of the QGP flow. One sees that the radial flow of the QGP pushes the dominance of coalescence over fragmentation to higher $p_\mathrm{T}$ region. With the presence of the medium flow, the coalescence contribution dominates the production of $D^0$ up to 5~GeV and $\Lambda_c$ up to 8~GeV, thus directly affecting the $\Lambda_c/D^0$ ratio up to 8~GeV. Our result is consistent with the findings presented in Refs.~\cite{Plumari:2017ntm,Cho:2019lxb} where a simplified parametrization of the medium flow was implemented. We  note that the so-called space-momentum correlation suggested in Ref.~\cite{He:2019vgs} is automatically included in our current and previous studies, because the local temperature and flow information of the surrounding fluid naturally enters the hadronization process of heavy quarks in our model.

\begin{figure}[tb]
  \epsfig{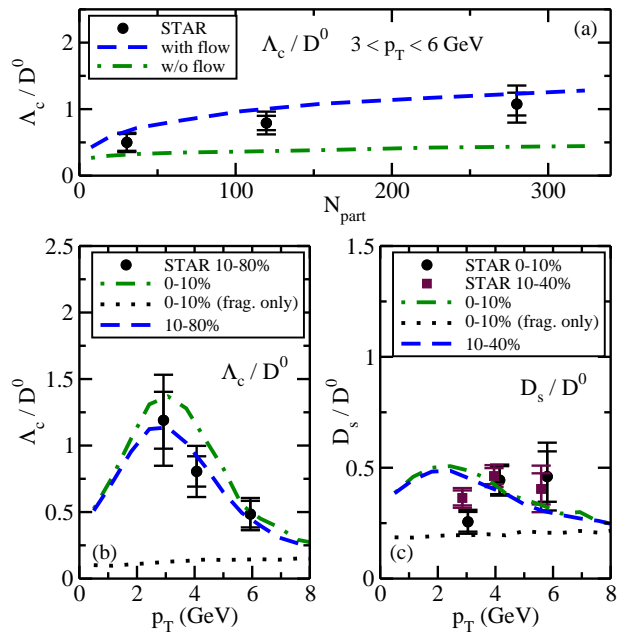}
  \caption{(Color online) Comparison of the $p_\mathrm{T}$-integrated $\Lambda_c/D^0$ ratio (a), the $p_\mathrm{T}$-differential $\Lambda_c/D^0$ ratio (b) and the $p_\mathrm{T}$-differential $D_s/D^0$ ratio (c) with the STAR data~\cite{Adam:2019hpq,Zhou:2017ikn} in 200~GeV Au-Au collisions.}
  \label{fig:plot-ratio_LD_DsD0}
\end{figure}

Figure~\ref{fig:plot-ratio_LD_DsD0} shows the $\Lambda_c/D^0$ and $D_s/D^0$ ratios at RHIC from our model. As shown in Fig.~\ref{fig:plot-ratio_LD_DsD0}(a), the $p_\mathrm{T}$ integrated ratio of $\Lambda_c/D^0$  (from 3 to 6~GeV) increases with the participant number of heavy-ion collisions due to the stronger radial flow of the QGP in more central collisions. This ratio is significantly reduced if the medium flow is switched off during the hadronization process, consistent with our finding in Fig.~\ref{fig:plot-spectra-AuAu200_0-10} that $\Lambda_c$ gains more transverse momentum from the medium flow than $D^0$ does due to its much larger mass. Shown in Fig.~\ref{fig:plot-ratio_LD_DsD0}(b) is the $p_\mathrm{T}$ differential ratio of $\Lambda_c/D^0$, which is greatly enhanced by the coalescence process as compared to the result from applying fragmentation alone. Our study also indicates that there is certain difference between results for the two centrality bins of 0-10\% and 10-80\%. Comparing the central collision results to the 10-80\% centrality data as in many theoretical studies in the literature is thus expected to give misleading conclusions.
We calculate the charmed hadron spectra in a wide centrality region (e.g. 10-80\%) according to $\langle dN_h/dp_\mathrm{T}\rangle=\Sigma_\mathrm{c} P^{(\mathrm{c})}dN_h^{(\mathrm{c})}/dp_\mathrm{T}$, where $P^{(\mathrm{c})} = N^{(\mathrm{c})}_\mathrm{bin}/\Sigma_\mathrm{c} N^{(\mathrm{c})}_\mathrm{bin}$ is the probability of finding heavy quark events in a given smaller centrality bin ``c", with $N^{(\mathrm{c})}_\mathrm{bin}$ being the binary collision number within this bin. Using only one average medium profile for the large centrality bin would lead to a much smaller nuclear modification effect. Figure~\ref{fig:plot-ratio_LD_DsD0}(c) shows the $D_s/D^0$ ratio, which is also significantly enhanced by the coalescence process as a result of the combined effects of the larger $D_s$ mass than $D^0$ and the enhanced strangeness production in a thermal medium compared to that in the vacuum fragmentation. We note that the result from using only the \textsc{Pythia} fragmentation deviates from the measured $\Lambda_c/D^0$ ratio in proton-proton collisions~\cite{Acharya:2017kfy} below $p_\mathrm{T}\sim 7$~GeV. And the default Peterson fragmentation function in \textsc{Pythia} does not distinguish between different charmed hadron species yet and also gives a steeper $D$ meson spectra in vacuum than expected. However, since the production of charmed hadrons in this $p_\mathrm{T}$ range from nucleus-nucleus collisions is dominated by the coalescence of charm quarks with quarks in the QGP as shown in Fig.~\ref{fig:plot-spectra-AuAu200_0-10}, the inaccuracy of \textsc{Pythia} should have a minor impact on the charmed hadron chemistry presented in this work. We will leave the fine tuning of the \textsc{Pythia} fragmentation to an upcoming effort.

\begin{figure}[tb]
  \epsfig{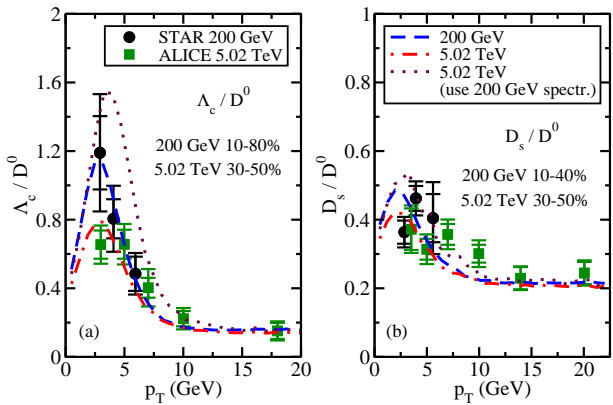}
  \caption{(Color online) Comparison of the $\Lambda_c/D^0$ and $D_s/D^0$ ratios between 200~GeV Au-Au collisions at RHIC~\cite{Adam:2019hpq} and 5.02~TeV Pb-Pb collisions at the LHC~\cite{Vermunt:2019ecg}.}
  \label{fig:plot-ratio_LHCvsRHIC}
\end{figure}

In Fig.~\ref{fig:plot-ratio_LHCvsRHIC}, we compare the $\Lambda_c/D^0$ and $D_s/D^0$ ratios between RHIC and the LHC. Since the radial flow of the QGP is much stronger in Pb+Pb collisions at the LHC than in Au+Au collisions at RHIC, one would naively expect to observe a larger value of $\Lambda_c/D^0$ ratio at the LHC than at RHIC, which is, however, opposite to what has been observed in experiments. This puzzling observation can, however, be naturally explained by the competition between the QGP flow and the charm quark $p_\mathrm{T}$ spectrum. As shown in Fig.~\ref{fig:plot-ratio_LHCvsRHIC}(a) by the dotted line, a much larger $\Lambda_c/D^0$ ratio would have been obtained if one uses the initial $p_\mathrm{T}$ spectra of charm quarks from RHIC for their evolution through the QGP medium produced at the LHC. On the other hand, the harder charm quark $p_\mathrm{T}$ spectrum at the LHC suppresses the QGP flow effect thus yields a smaller  $\Lambda_c/D^0$ ratio. We observe in Fig.~\ref{fig:plot-ratio_LHCvsRHIC}(b) a similar but smaller effect for the $D_s/D^0$ ratio. 

{\em Summary. --}
We have developed a comprehensive coalescence-fragmentation approach for studying heavy quark hadronization in heavy-ion collisions. Both $s$ and $p$-wave charmed hadron states, which are sufficient to cover all major charmed hadron states reported in PDG, have been included in our coalescence model. We have found that the inclusion of $p$-wave states enhances the total coalescence probability of charm quarks. It is also necessary for a full 3-dimensional calculation to normalize the charm-QGP coalescence probability at zero momentum using reasonable in-medium sizes for charmed hadrons. In our new coalescence model, a strict 4-momentum conservation has been implemented by first forming the off-shell excited hadron states and then letting them decay into the ground state charmed hadrons, which guarantees the boost invariance of the coalescence probabilities for producing charmed hadrons and the thermal limit of their energy spectra. By combining our new hadronization approach with the up-to-date FONLL+EPPS16 initial spectra of charm quarks and their nuclear modification through the advanced Langevin-hydrodynamics model, our state-of-the-art calculation has provided a simultaneous description of the $p_\mathrm{T}$-integrated and differential $\Lambda_c/D^0$ and $D_s/D^0$ ratios at both RHIC and the LHC. We have also found that the interplay between the QGP flow and the charm quark transverse momentum spectrum is essential for describing the final charmed hadron chemistry in heavy-ion collisions, especially the puzzling observation of a larger $\Lambda_c/D^0$ ratio at RHIC than at the LHC. Our study has further suggested that the sizes of charmed hadrons should be larger in medium than in vacuum, which is qualitatively consistent with the findings in Ref.~\cite{Shi:2019tji} and may be further tested by hadronic model calculations in the future.

{\em Acknowledgments --}
We are grateful to helpful discussions with Weiyao Ke, Rainer Fries and Shuang Li. This work was supported in part by the U.S. Department of Energy (DOE) under Grants No. DE-SC0013460 and DE-SC0015266, the National Science Foundation (NSF) under Grant No. ACI-1550300, the Welch Foundation under Grant No. A-1358, the Natural Science Foundation of China (NSFC) under Grants No. 11775095, 11805082, 11890711, and 11935007, and by China Scholarship Council (CSC) under Grant No. 201906775042.

\bibliographystyle{h-physrev5}
\bibliography{SCrefs}

\end{document}